\documentclass[twocolumn,aps,pra,showpacs]{revtex4}
 
\usepackage{epsfig}
\usepackage{amsmath}
 
\begin{document}
\renewcommand{\theparagraph}{\Alph{paragraph}}
  
\title{Recoil-induced lasing}

\author{J.~Javaloyes$^1$, M. Perrin$^1$, G.L. Lippi$^1$, and A. Politi$^{1,2}$}

\affiliation{
$^{1}$
Institut Non Lin\'eaire de Nice, UMR 6618 CNRS, Universit\'e de
Nice-Sophia Antipolis, 1361 Route des Lucioles,\\
F-06560, Valbonne France \\
$^{2}$
Istituto Nazionale di Ottica Applicata, L.go E. Fermi 6, 50125 Firenze, Italy\\
}
\date{\today}
\begin{abstract}

The interaction of an atomic gas confined inside a cavity with a strong
electromagnetic field is numerically and theoretically investigated in a regime
where recoil effects are not negligible. The spontaneous appearance of a
density grating (atomic bunching) accompanied by the onset of coherent
backpropagation is found to be ruled by a continuous phase-transition.
Numerical tests allow us to convincingly prove that the transition is steered
by the  appearence of a periodic atomic density modulation.  Consideration of
different experimental relaxation mechanisms induces us to analyze the problem
in nearly analytic form, in the large detuning limit, using both a Vlasov
approach and a Fokker-Planck description.  The application of our predictions
to recent  experimental findings, reported in [Phys. Rev. Lett. {\bf 91},
183601 (2003)], yields a semi-quantitative agreement with the observations.

\end{abstract} \vspace{2mm}
\maketitle

%
%
%%%%%%%%%%%%%%%%%%%%%%%%%%%%%%%%%%%%%%%%%%%%%%%%%%%%%%%%%%%%%%%%%%%%%%%%%%%%%%
%                               MAIN TEXT
%%%%%%%%%%%%%%%%%%%%%%%%%%%%%%%%%%%%%%%%%%%%%%%%%%%%%%%%%%%%%%%%%%%%%%%%%%%%%%

\section{Introduction}\label{intro}

Recoil in the interaction between atoms and electromagnetic fields is
almost exclusively associated with the idea that momentum transfer is a
way of slowing down atoms to extremely low-temperatures \cite{cool}.
However, non trivial consequences on light propagation have been uncovered as
well. For instance, in the realm of single-atom effects, the so-called
Recoil Induced Resonances (RIR) have been investigated both
theoretically~\cite{GBD92} and experimentally~\cite{CGLV92}. Moreover,
it has been conjectured that atomic recoil can collectively give rise to
coherent propagation through backreflection from a spontaneously generated
density grating: the so called Collective Atomic Recoil Laser (CARL)~\cite{carl}.
Finally, connections between the two approaches have been investigated
in~\cite{Berman99}.

The very first observation of spontaneous backward emission was made in a ring
cavity experiment~\cite{OKRL89}, where the very low ``transverse temperature"
might have been responsible for CARL amplification. More detailed experimental
accounts have been reported in Refs.~\cite{LBBT96,HBK96},
although it was not possible to establish whether the generation of the
backward field was due to the spontaneous formation of a density grating, or
viceversa. In fact, although it was later ascertained that recoil plays a
prominent  role~\cite{BFL99} in the underlying physics~\cite{LBBT96,HBK96}, an
alternative interpretation based on the formation of a standard polarization
grating could not be entirely ruled out~\cite{BGGV97}. Only recently has the
first convincing evidence of CARL  been given in a beautiful experiment
performed on a sample of cold Rubidium atoms~\cite{KCZC03}.

On the theoretical side, too, no final conclusion could be drawn on the CARL
features, because the first model allowed only investigating the transient
regime. Stationary states have been obtained for the first time by including
collisions with an external buffer gas \cite{PLP01}. By simulating a
low-temperature (a few mK) sodium  vapour in the detuning range previously
considered in the (numerical) literature, a nonequilibrium phase transition
leading to a stationary nonzero backward field (above a given threshold) was
therein identified. However, such a collective phenomenon could not be linked
to the onset of a density grating, but rather to the birth of a nontrivial
polarization grating~\cite{PLP01}.

The overall scenario has been observed also in a more general model accounting
for the input-field dynamics~\cite{PYN02} and has been successively clarified
in~\cite{JLP03} with the help of the elimination of the atomic variables;
indeed, this step rendered the derivation of an effective free-energy potential
for the backward field amplitude possible, thereby clearly showing the
existence of  a phase transition.

In this paper we show numerically and theoretically that this latter model also
accounts for the CARL and allows one to understand its onset in terms of
a continuous phase transition characterized by two effective variables. The
resulting estimate of the critical point is compatible with the experimental
results reported in \cite{KCZC03}. The results presented here provide the first
substantial and systematic progress towards a proof of the existence, in stable
form, of a collective interaction, which, predicted years ago~\cite{carl}, had
so far proven to be elusive.

More precisely, we investigate the transition which gives rise, spontaneously
-- i.e., without an external probe field --, to a coherent electromagnetic
field counterpropagating with respect to the injected one. We choose to consider
the ring-cavity-based model introduced in \cite{YN01,PYN02} for two reasons. On
the one hand, the available single-pass model~\cite{carl} has intrinsic
shortcomings~\cite{com1} which may be of
little consequence in particular circumstances, but which renders it
nonetheless less attractive. On the other hand, the most recent experimental
results~\cite{KCZC03}, which can be considered at the present time as the first
and only available observation of a true CARL system, have been obtained in a
ring cavity.  Hence, the choice of a bidirectional ring cavity model imposes
itself.

A strong concern, which tainted all previous predictions and experimental
attempts at studying this problem, regards the influence of external
perturbations (collisions, finite transit time in the interaction volume,
``optical friction'', etc.) on the collective dynamics. One can conjecture that
their presence could destroy the phase transition or, at least, affect it so
strongly as to shift it to an unphysical region of parameter values. Our
purpose in this work is to investigate such mechanisms using different
approaches and mathematical/numerical tools, to obtain a convincing picture.
Section~\ref{Mod} introduces the model we used in the investigation and
discusses, in subsection\ref{relax}, the main physical relaxation mechanisms that
may take place in an experiment. The appropriate modeling tools for analyzing
the two main mechanisms are introduced here. Section~\ref{num} is devoted to a
systematic numerical analysis which allows us to highlight the origin of the
forces that give rise to a transition of the CARL-type -- i.e., caused by the
appearence of a spatial density grating -- as opposed to the one recently
predicted~\cite{PLP01,JLP03}, based on an unusual polarization grating and
possibly responsible for high-temperature experimental
observations~\cite{LBBT96,HBK96}. The numerical approach makes it possible to
isolate the different source terms and identify in which range of temperature
each of them is predominant.  Analytical descriptions present the advantage of
offering a better understanding of the physics driving the system. Since the
full problem is obviously analytically unmanageable, it is useful to study some
limiting situations.  This is done in Section~\ref{Large}, where we consider the
case of large detuning between fields and atomic resonance. In this spirit, we
describe the perturbations introduced by transit broadening
(subsection~\ref{Vlasov}) and show that a second-order transition appears
which can be described by expanding the spatial inhomogeneities in Fourier
modes. The first and most relevant mode coincides with the {\it bunching
parameter}, previously introduced to quantify the density grating~\cite{carl}.
Losses introduced by an optical molasses are studied in subsection~\ref{FP}
with the help of a Fokker-Planck equation that can be analytically solved in
the large dissipation limit. Predictions resulting from the preceding analyses
are offered in Section~\ref{predictions}, together with a comparison with the recent
experimental results~\cite{KCZC03}. Some conclusions and perspectives are
presented in Section~\ref{concl}.

\section{Model}\label{Mod}

The starting model involves four dimensionless variables (one being complex)
accounting for
the single atom dynamics plus two complex equations describing the
amplitude of forward ($x_f$) and backward ($x_b$) uniform fields~\cite{PYN02}
\begin{eqnarray}
\dot \theta_j &=& p_j , \nonumber \\
\dot p_j &=& - \ \Re e \!\left[ ( x_f {\rm e}^{i\theta_j} -
            x_b {\rm e}^{-i\theta_j}) s_j^* \right](|\Delta_a|/2G), \nonumber \\
\dot s_j &=&   G (x_f {\rm e}^{i\theta_j} +
 x_b {\rm e}^{-i\theta_j})d_j  - (\Gamma_\perp \!-\!i\Delta_a)s_j,
 \label{general} \\
\dot d_j &=& - G \Re e \!\left[ (x_f{\rm e}^{i\theta_j} +
    x_b{\rm e}^{-i\theta_j})s_j^*\right] - \Gamma_\parallel(d_j+1), \nonumber\\
\dot x_f &=& -(1+i \Delta_c)x_f + Y +
      {\tilde C} \left\langle s{\rm e}^{-i\theta}\right \rangle, \nonumber\\
\dot x_b &=& -(1+i \Delta_c)x_b +
   {\tilde C} \left\langle s{\rm e}^{i\theta}\right\rangle ,  \nonumber
\end{eqnarray}
where $1\le j \le N$ and the angular brackets denote an average over the atomic
ensemble. Time and related parameters are expressed in units of the photon life-time
inside the cavity, $1/\kappa$, namely: the decay rates of the
atomic polarization ($\Gamma_\perp$) and population inversion
($\Gamma_\parallel$), the detuning  $\Delta_a$ between input field and atomic
frequencies, and  the cavity detuning $\Delta_c$ (i.e. the distance from the
nearest cavity resonance). Moreover, $G = \sqrt{m\kappa|\Delta_a|/\hbar}/k$ and
$\tilde C=\alpha \Gamma_\perp L/(G{\mathcal T})$ are the two coupling
constants ($m$ is the atomic mass, $k$ the field wavenumber, $\alpha$ the
unsaturated absorption rate per unit length, $L$ the length of the atomic
medium, and ${\mathcal T}$ the  cavity's transmittivity).
The amplitude $Y$ of the injected field is scaled such that
$Y^2=4 \mathcal{D}^2 {\mathcal P}/(AG^2 {\mathcal T} \kappa^2 \hbar^2)$, where
$\mathcal{D}$ is the atomic dipole moment, ${\mathcal P}$ is the input power, and $A$ the beam
surface, expressed in physical units. An equivalent scaling is used for
the forward and backward intracavity components, $x_f$ and $x_b$.
Finally: (i) $\theta_j = k z_j$ is the scaled atomic position;
(ii) the atomic momentum $P_j$ is rescaled to $p_j = k P_j/(m \kappa)$;
(iii) $s_j$ and $d_j$ are the atomic polarisation and population
inversion, the first being expressed in units of $\mathcal{D}$.

\subsection{Relaxation mechanisms}\label{relax}

While dissipation mechanisms affecting the internal degrees of freedom and the
field amplitudes are accounted for by the above model, no losses acting on the
atomic momentum are therein included. As a result, atoms are continuously
accelerated by radiation pressure, and Eqs.~(\ref{general}) can thereby describe
only the transient regime corresponding to the early stages of CARL amplification.

In actual experimental situations, relaxation processes may be induced
by various mechanisms, the main ones being: (a) collisions with a buffer gas (those
between optically active atoms can be neglected under all realistic
circumstances), (b) finite residence time in the interaction volume (added to some
relaxation occurring outside it), and (c) ``viscous'' losses (e.g., as in an
optical molasses). Collisions with a buffer gas require fairly high pressures
and have been often used in the investigation of nonlinear coupling between
atoms and light (cf., e.g.,~\cite{MDLM86,AL01}).
In a numerical scheme, this relaxation processes can be implemented as a
coupling with a thermal bath, efficiently simulated with ``molecular dynamics''~\cite{MDT78}
techniques.
More precisely, a random sequence of inter-collision times is independently generated for each
atom according to an exponential distribution with average value equal to
$1/\gamma_c$. At each collision, the momentum of the colliding, $j$-th, atom is
randomly reset according to the Gaussian distribution
$Q_{0}(p_{j})=\exp\left\{-p^{2}_{j}/(2T)\right\}/\sqrt{2\pi T}$
where $T=k_{B}\bar{T}k^2/(m \kappa^2)$ is the rescaled temperature of the
buffer gas ($k_{B}$ is Boltzmann's constant, $\bar{T}$ is the temperature
in physical units); moreover, the phase of the
atomic polarization $s_{j}$ is also reset to a value uniformly distributed in
the whole range $[0,2\pi]$, as $|s_j|$ remains unchanged. Notice that this
algorithm amounts to implicitely assuming that active and buffer-gas atoms have
equal masses.

The finite transit time (b) introduces effective losses and has certainly
played the main role in several experiments~\cite{OKRL89,LBBT96,HBK96}. In such a
situation, the atoms spend a more or less small fraction of time inside the
interaction volume, which they enter and exit at random positions and in a
random state (both for the velocity and for the internal degrees of freedom).
Outside the interaction region, the atoms suffer thermalization, either through
collisions with other atoms (e.g., a low pressure  buffer gas), with a
container, or are simply renewed through ``fresh'' atoms coming from a beam;
hence, the effective relaxation comes from the loss of ``synchronized'' atoms
(both for the external and the internal degrees of freedom) compensated, in
average, by the
entrance into the interaction volume, of ``random'' atoms. This phenomenon can
be reinterpreted as a ``reset'' in the external and internal degrees of freedom
of the atom in question. At the microscopic level, one can still adopt a 
description based on the thermalization scheme discussed in (a), by modelling
each random exit/entrance in the active volume as a ``collision"; 
the main difference with the previous case is that all atomic variables have
now to be reset. At the macroscopic level, the mathematical tool
most suited for such a description is a Vlasov equation, where the
probability distribution for the atomic position and momentum are dynamically
investigated. These considerations are the basis for the discussion
of Section~\ref{Vlasov}.

The third mechanism (c), recently employed in~\cite{KCZC03}, is an optical
molasses, added in the experiment to obtain steady state backward emission.
Here, the ``molasses field'' slows down the atoms by the standard cooling
action while introducing a ``friction'' which prevents the Doppler shift from
growing so large as to render the interaction with the field entirely
negligible. It has been shown~\cite{Cohen} that under the assumption of a small
saturation parameter of the molasses field, the momentum dynamics is well
described by a Langevin equation (compare with the second of
Eqs.~(\ref{general})),
\begin{equation}
\dot p_{j} =  \Re e \!\left[ (x_b {\rm e}^{-i\theta_{j}} -
      x_f {\rm e}^{i\theta_{j}}) s_{j}^* \right](|\Delta_a|/2G) -\gamma p_{j} +
	     \eta_{j},
\end{equation}
where $\eta_{j}$ is a standard white noise,
\begin{eqnarray}
\label{white_noise}
\langle \eta_{j}(t) \rangle &=& 0, \\
\langle \eta_{j}(t_1) \eta_{k}(t_2) \rangle &=& 2 \gamma T \delta_{jk}\delta(t_1-t_2), \nonumber
\end{eqnarray}
 ($\delta_{jk}$ is the Kronecker symbol)
accounting for the inevitable fluctuations that thermalize the atom to a finite
temperature $T$. In Section~\ref{FP}, we will study this model in the limit of
large detuning, a condition well verified in the experiment~\cite{KCZC03}.

\section{Numerical Results}\label{num}

We now intend to explore the predictions of the model, Eqs.~\ref{general}, from a
purely numerical point of view to provide a starting point for the later
analysis.  In particular, we are interested in studying the existence of a
{\it steady state bunching} in the presence of a backward propagating field.
We remark that the observation of a stable value of bunching in a bidirectional
ring cavity has never been reported. Previous observations, obtained without
cavity~\cite{carl}, have shown a strongly oscillating,
irregular behaviour for this variable. In the other kind of observed transition
~\cite{PLP01,PYN02}, instead, the $\theta$ distribution remains flat
(within the fluctuations imposed by the sample size) both below and above
threshold.  Hence, the mere numerical observation of a stable, steady bunching
represents an important step towards generalizing the CARL
predictions~\cite{carl}.

Such an observation is, however, not sufficient in itself. The question
arises naturally, whether the backward field is originated by the growth of a
nonhomogenous longitudinal atomic distribution, or whether the spatial
inhomogeneity results as a consequence of a preexisting optical standing wave,
generated by other source terms. Although apparently obvious, this point is
far from being trivial. Indeed, any standing wave is going to generate a
whole hierarchy of gratings (atomic polarization and inversion, index of
refraction and -- if the atoms are free to move -- atomic density), but one
needs to identify the true source.  In this system, a source term which gives
rise to a collective state has already been identified in the nontrivial
grating in the atomic polarization's phase~\cite{PLP01,JLP03}, while a
standard polarization modulation (without phase transition) has been proposed~\cite{BGGV97}
as a possible mechanism for interpreting experimental observations~\cite{LBBT96,HBK96}.

\begin{figure}[ht!]
\includegraphics[width=8.cm]{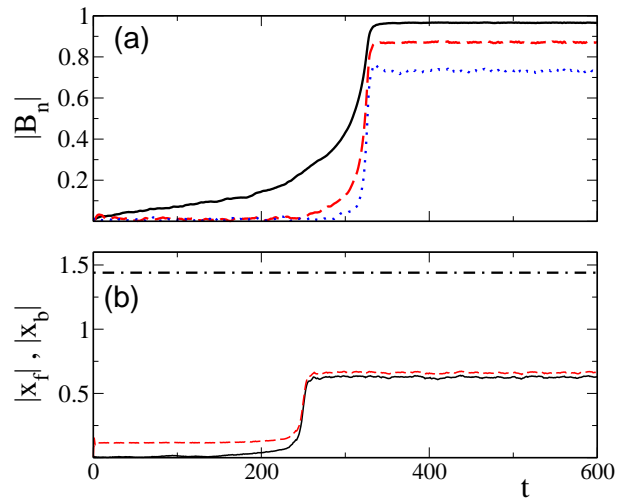}
\caption{Fig.~(a), shows the time evolution of amplitude of the first Fourier
modes of the $\theta$-distribution ($n=1,2,3$ correspond to solid, dashed and
dotted curves, respectively).  Fig.~(b) shows the rise of both amplitudes
(solid and dashed line corresponds to $|x_b|$ and $|x_f|$). The dot-dashed line
represents the injected field, $Y$. Parameter are: $\Delta_a =-30$,
$\Gamma_\perp=1$,  $\Gamma_\parallel=2$, $\tilde C=7.1$, $\Delta_c=0$, $G=54$,
$Y=1.45$, $\gamma_c=1/3$, $T =5.10^{-2}$.}
\label{Fig_rise}
\end{figure}

In order to characterize the appearence of a spatial inhomogeneity in the
atomic distribution, we introduce the characteristic function of the position
($z_j$) distribution defined in terms of its Fourier modes
\begin{equation}
\label{Bunch_def_general}
B_n := \frac{1}{N} \sum_{j=1}^N  {\rm e}^{2in k z_j}.
\end{equation}
Introducing the normalised variable $\theta$, the general definition
Eq.~(\ref{Bunch_def_general}) reads, with our normalisation,
\begin{equation}
\label{Bunch_def}
B_n = \langle {\rm e}^{2i\text{n}\theta} \rangle.
\end{equation}
The direct integration of model (\ref{general}), using the same techniques as
outlined in~\cite{PLP01} (i.e. with thermalization induced by collisions),
provides the results shown in Fig.~\ref{Fig_rise}. Fig.~\ref{Fig_rise}a shows the
appearence of several Fourier modes, which reach a steady
state amplitude after the backward field has grown to its final value
(Fig.~\ref{Fig_rise}b, solid line). We stress that the latter grows from zero
spontaneously, without any {\it seed} being injected in the computation.

From Fig.~\ref{Fig_rise}a we remark that one Fourier component of the atomic
spatial distribution significantly grows away from zero well before the
backward field begins to deviate from its initial value. We will take this as
a suggestion, to be later verified, that the bunching is responsible for the
appearence of the counterpropagating field.
The harmonic components, instead, seem to be generated by the interaction of
the inhomogeneously distributed atoms and the optical standing wave generated
by the two macroscopic counterpropagating fields, ($x_f$ and $x_b$).
In addition, we also see how the forward field (dashed line in
Fig.~\ref{Fig_rise}b) is modified by the presence of the spatially modulated
atomic distribution.

The physical picture described by this figure is the following: the cavity initially
contains the atoms and no fields. At time $t=0$, we start injecting the external
component $Y$, collinear with the forward direction.
During the transient preceding synchronisation, the atoms use part of the energy
to adapt their positions and velocities, thus building up a non homogeneous spatial
distribution.
We remark that the total energy injected into the cavity (see dot-dashed line)
is larger than that contained in the two field components at steady
state ($t > 300$ time units). The fact that both fields, $x_f$ and $x_b$,
grow nearly to the same value implies that one cannot neglect the dynamics of
the forward field. This dynamics maximizes the spatial modulation and {\it
both} fields adapt themselves to the global coupling that ensues. It is also
crucial to remark that, although the spatial atomic inhomogeneity appears to
be responsible for the generation of the backward field, at time $t \approx
230$ time units, the $B_1$ component has reached only one third of its final
value. Hence, the rest of the modulation (fundamental plus all the harmonics),
results from the interaction with the optical standing wave. This point is
significant, since it implies that the scattering from one field into the
other is going to be quite symmetric: any increase in one component will imply
a larger number of photons in one direction, hence a larger scattered
contribution from that direction into the other. The residual difference
between the forward (dashed line, Fig.~\ref{Fig_rise}a) and the backward field is
to be attributed to the fact that the former receives an additional
contribution from the external injection; hence the corresponding mode
contains more photons at any time.
\bigskip

\begin{figure}[ht!]
\includegraphics[clip,width=6.5cm]{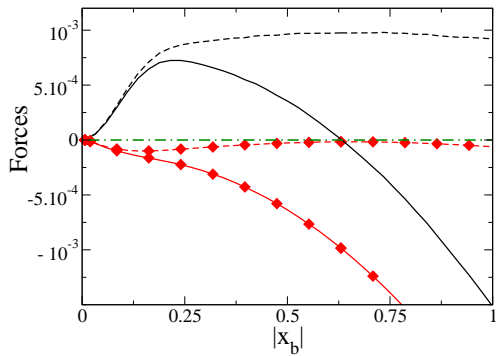}
\caption{The force $\mathcal{F}_{\text{A}}$ (dashed lines) and
$\mathcal{F}$ (solid lines) versus the backward field amplitude $x_b$.
Curves with no symbols correspond to the simulation of the full model;
diamonds display curves obtained by imposing a flat distribution of positions
$\theta$ (same parameter values as in Fig.~\ref{Fig_rise}).}
\label{forcefield}
\end{figure}

A more convincing illustration of the active role played by the density grating
in the generation of the backward field, can be gained by decomposing the
``force" field acting on the intensity $|x_b|^2$ of the backward field into two
contributions
\begin{equation}
\frac{d|x_b|^2}{dt} = \mathcal{F}(x_f,x_B) = -2|x_b|^2 +
\mathcal{F}_{\text{A}}(x_f,x_b),
\label{force1}
\end{equation}
where $\mathcal{F}_{\text{A}}(x_f,x_b) = 2 \tilde C
{\Re e}\left(\langle s\text{e}^{i\theta}\rangle x_b^* \right)$. The first term
on the r.h.s., due to cavity losses, has always a stabilizing effect, while
$\mathcal{F}_{\text{A}}(x_f,x_b)$, which accounts for the atomic contribution,
can be either stabilizing or destabilizing. In order to gain some insight on the
resulting dynamics, we compare the force fields generated in the presence and
in the absence of bunching, respectively. We do this under the assumption
that the atomic degrees of freedom can be neglected~\cite{note1}, i.e. that the
relevant dynamical properties are contained in the dependence of
$\mathcal{F}_{\text{A}}$ on $x_b$ and $x_f$. As long as the atomic variables
rapidly converge to a stationary state, one can treat $|x_f|$ and $|x_b|$ as if
they were fixed parameters and thereby numerically determine the ``force"
field for different choices of the two amplitudes. In principle, one should also
take care of the time-dependence associated with the detuning $\nu$ between the
two fields; however, numerical simulations of the full model have revealed
that the ``force" field
depends very weakly on $\nu$. The force resulting for the same parameter
values as in Fig.~\ref{Fig_rise} is plotted in Fig.~\ref{forcefield} for
different values of the backward field amplitude. The solid line with no
symbols represents the total effective ``force'' acting on the backward field.
Steady state operation exists only when the resulting total force is zero,
and its stability is obviously determined by the local slope. It is clear
that the zero-field state is unstable, while $|x_b| \approx 0.6$ is a
stable one. In this latter regime, one can state that the destabilizing action
of the atomic contribution, ${\mathcal F}_{\text{A}}$ (dashed curve without
symbols in Fig.~\ref{forcefield}), is balanced by the stabilizing effect due to
the cavity loss term, $|x_b|^2$.

In order to test of the {\it true role} played by atomic bunching (i.e., as a
{\it master} or a {\it slave} variable), we have proceeded to ``switching off"
both collisions and optical forces acting on the atoms (this is obviously
possible only in a simulation). The corresponding curves representing
${\mathcal F}_{\text{A}}$ (dashed line in Fig.~\ref{forcefield}) and the total
force, ${\mathcal F}$, (solid line) are identified by symbols. As a result of
the absence of external perturbations, each atomic position $\theta_j$ evolves
linearly according to the velocity value at the ``switch off" time; in this
way, the density grating is rapidly washed out. This procedure allows us to
separate the possible gain component coming from the density grating from the
one arising in a pumped ensemble of two-level systems \cite{HH70}. The
stationary value of $\mathcal{F}_{\text{A}}$ under such conditions remains
always negative (so does {\it a fortiori} the total force), indicating that the
presence of a density grating is an {\it indispensable} ingredient to maintain a
finite backward field. Notice that this is at variance with the transition
studied in \cite{PLP01}, where the force field would be positive even in the
absence of a density grating \cite{JLP03}. Hence, we can safely conclude that
the onset of a finite backward field is a clear example of CARL and that the
steady state bunching obtained is the {\it cause} for the appearence for the
backward field component, and not {\it a consequence} of the existence of an
optical standing wave (cf., e.g.,~\cite{BGGV97}) in the parameter range which
we are investigating.

\section{Large detuning limit}\label{Large}

In Section~\ref{num} we have just reported purely numerical observations. In
the perspective of constructing an analytical description of the transition,
it is important to identify the possibly few relevant variables. For this
purpose we concentrate on the large detuning limit, where the standard
adiabatic elimination holds. Although this allows us to analytically treat
only a limiting case, we will draw from it the benefit of later comparing the
predictions of this approach to the recent experimental
observations~\cite{KCZC03}, obtained with very large detuning between field
and atoms ($\Delta_a \approx 10^6 \gg 1$).  A discussion of the predictions
resulting from such an analysis will be offered in Section~\ref{predictions}.

By setting, in Eqs.~(\ref{general}), $\dot d_j = \dot s_j = 0$, expanding the
solutions in powers of $1/\Delta_a$ (to be considered as a smallness parameter)
and considering first order terms, one finds $d_j = -1$ and
\begin{equation}
\label{polariz}
s_j = -\frac{iG}{\Delta_a}
(x_f {\rm e}^{i\theta_j} + x_b{\rm e}^{-i \theta_j}) \quad.
\end{equation}
Accordingly, our model, Eqs.~(\ref{general}), reduces to the following set of
equations
\begin{eqnarray}
\dot{\theta}_{j}  &=& p_{j}, \nonumber \\
\label{reduced}
\dot{p}_{j}  &=&\Im m(x_f x_b^{\ast}{\rm e}^{2i\theta_j}) ,\\
\dot{x}_f  &=&  -\left[(1+ i {\bar \Delta}) x_f - Y + i C x_b B_{-1}\right] ,
\nonumber \\
\dot{x}_b  &=& - \left[ (1+ i{\bar \Delta}) x_b + i C x_f B_{1}\right]
\nonumber ,
\end{eqnarray}
where $ C =\tilde C G/\Delta_a$, and ${\bar \Delta} = \Delta_c + C$ accounts
for the combined effect of atomic and cavity frequency shifts in the absence of
bunching (i.e. $B_1=B_{-1}=0$). The above form is similar to the one discussed
in \cite{BV96}, which was, however, obtained by phenomenologically adding
relaxation terms to a single pass model~\cite{B2}.

\subsection{Vlasov Model}\label{Vlasov}

It is now convenient to pass to a Vlasov-like description by introducing the
distribution $Q(\theta,p)$ of positions and momenta. In order to simplify the
analysis we now assume that also the atomic positions are randomized in each
collision~\cite{note3}. Accordingly, we can write
\begin{equation}
\partial_{t}Q + p\partial_{\theta}Q + \Im m(x_f x_b^{\ast} {\rm e}^{2i\theta})
\partial_{p}Q = -\gamma_c(Q-Q_0) ,
\label{eqvlasov}
\end{equation}
where $\partial_y$ denotes the partial derivative with respect to $y$.
In the absence of a backward field, $Q(\theta,p)$ converges towards the
equilibrium distribution $Q_0$.

We now study the stability of the solution $x_f =Y/(1+i\bar \Delta):=
Y_0$, $x_b=0$, and $Q = Q_0$, by introducing the perturbations $\delta x_f$,
$\delta x_b$, $\delta Q$.
Since the equation for $x_f$ depends on the other variables only at second
order in the perturbations, it can be solved separately. Moreover, $\delta x_f$
affects $\delta x_b$ only at second order, so that it can be factored out.
Therefore, the stability of $\delta x_f$ justifies the assumptions made in
deriving the single-pass model where $x_f$ is considered to be a constant
parameter~\cite{carl,PLP01,JLP03}. One is thus left with the following two
equations
\begin{eqnarray}
\label{eqself_a}
\dot{\delta x_b}  &=& - (1+i\bar \Delta)\delta x_b - iC \bar Y \delta B_{1} , \\
\label{eqself_b}
\partial_{t}\delta Q &=&  - \partial_{p}Q_0 \Im m (\bar Y \delta x_b^{\ast}
{\rm e}^{2i\theta})
-(\gamma_c + p\partial_{\theta})\delta Q .
\end{eqnarray}
From the structure of such equations we see that if the bunching is washed out --
i.e., $B_{\pm 1} = 0$ and $\delta Q = 0$ -- no instability can be expected in
this regime. This is at variance with the case discussed in \cite{PLP01,JLP03}
where the effective stability of $x_b$ can be turned into an instability
sufficiently close to the atomic resonance.

From Eq.~(\ref{eqself_a}), one realizes that the coupling occurs through the
first Fourier mode of the density distribution. One can thus solve
the above equation by introducting the following Ans\"atze,
\begin{equation}
  \delta Q(\theta,p,t) = {\cal Q}(p) {\rm e}^{-2i\theta + \lambda t} + c.c.
  \quad , \quad \delta x_b = {\cal E}_b {\rm e}^{\lambda t} ,
\label{ansatz}
\end{equation}
where $\lambda$ is a complex stability eigenvalue. After substituting
the above expressions into Eqs.~(\ref{eqself_a}, \ref{eqself_b}) and equating
terms with the same exponential factor, one finds
\begin{eqnarray}
\label{Vlasov1}
 \left(\lambda + 1 + i\bar \Delta \right) {\cal E}_b  +
2i \pi C \bar Y \int_{-\infty}^{\infty} dp  {\cal Q}(p)  &=& 0,  \\
\label{Vlasov2}
 \frac{i}{2} \bar{Y}^{\ast} {\cal E}_b \partial_p Q_0 +
  (\lambda - 2ip + \gamma_c) {\cal Q}  &=& 0.
\label{eqauto}
\end{eqnarray}
By solving Eq.~(\ref{Vlasov2}) and replacing into Eq.~(\ref{Vlasov1}), one obtains
the solvability condition
\begin{equation}
\label{lambda}
 \lambda = -\left(1 + i\bar \Delta\right)  -\pi C |\bar Y|^2 \int_{-\infty}^{\infty}
   dp \frac{\partial_p Q_0}
 {\lambda - 2ip + \gamma_c}.
\end{equation}
Within this framework, the onset of both bunching and backward field is
signalled by a change of sign of the real part of $\lambda$,
$\lambda_r$\cite{note4}. The integral in Eq.~(\ref{lambda}) can be analytically
expressed using an error function. Nevertheless, this would hide an important
point. Let place ourselves at threshold, and rewrite  Eq.~(\ref{lambda}),
assuming that $\lambda$ is purely imaginary. Setting  $\lambda = -i\beta$, we
obtain for the real part
\begin{equation}
\label{lambda_real}
 0 = -1  + \pi \frac {C \gamma_c |\bar Y|^2}{T \sqrt{2\pi T}}
\int_{-\infty}^{\infty} dp \frac{p~{\rm e}^{-p^2/(2T)}}{(\beta + 2p)^2 +
\gamma_c^2}.
\end{equation}
In order for such an equation to be valid, the condition
\begin{equation}
\label{lambda_cond}
\int_{-\infty}^{\infty} dp \frac{ p~{\rm e}^{-p^2/(2T)}}{(\beta + 2p)^2 +
\gamma_c^2} > 0
\end{equation}
must be fulfilled. The numerator is an odd function of $p$. The sign of the
integral in  Eq.~(\ref{lambda_cond}) is thus determined by the denominator,
whose minimum is reached at $p_o=-\beta/2$. Finally, Eq.~(\ref{lambda_cond})
imposes $\beta < 0$. From the sign convention adopted here (the same that in
Ref.~\cite{PYN02}), a complex field $E$, of frequency $\omega$, is linked to its
slowly varying envelope $A$ as $E=A\exp [-i \omega t]$. Thus $\beta$ physically
represents the detuning between the backward field and the forward field.
One can thus conclude that the spontaneous symmetry breaking causing $x_b$ to
rise can only give rise to a red detuned field.

\subsection{Fokker-Planck description}\label{FP}

We now turn to analyzing the relaxation mechanism that most closely
models the recent experimental results~\cite{KCZC03}. There, the cw
superposition of additional laser beams, tuned to the D$_2$ transition of
Rubidium, introduces a mechanism which slows down those atoms that would be
accelerated away from resonance by the strong interaction arising from CARL.
Indeed, it was reported~\cite{KCZC03} that in the absence of the molasses only
transient backwards emission could take place (and its interpretation as CARL
is complicated by the technique used for preparing the sample, which pre-traps
the atoms in a standing wave). The key point in the experiment is that the
molasses is always present during the interaction but on the {\it other}
D-line (the experiment being done on the D$_1$).

In the large detuning limit, the second of Eq.~(\ref{reduced}) is replaced by
\begin{equation}
\label{FP_extern}
\dot{p}_{j} =\Im m(x_f x_b^{\ast}{\rm e}^{2i\theta_{j}}) -\gamma p_{j} + \eta_{j}(t).
\end{equation}
Eq.~(\ref{FP_extern}) describes the evolution of ($\theta$, $p$) in a potential
self-consistently determined by the dynamics of the two counter-propagating
field components, $x_f$ and $x_b$, which in turn depends on the joint distribution
$Q(\theta,p)$. The evolution of $Q$ corresponding to Eq.~(\ref{FP_extern}),
is given by the Fokker-Planck equation
\begin{eqnarray}
\label{evol_Q_general}
&& \partial_{t}Q + p\partial_{\theta}Q + \Im m(x_f x_b^{\ast}{\rm e}^{2i\theta})
\partial_p Q = \\ \nonumber
&& \gamma \partial_p \left( pQ + T \partial_p Q\right) .
\end{eqnarray}
Notice that this equation differs from the Vlasov equation (\ref{eqvlasov})
only for the relaxation terms contained in the right hand side.

By proceeding as in the previous subsection, we linearize
Eq.~(\ref{evol_Q_general}) obtaining
\begin{eqnarray}
\label{eqself_FP_b}
\partial_{t}\delta Q =  &-& \partial_{p}Q_0 \Im m (\bar Y \delta x_b^{\ast}
{\rm e}^{2i\theta})\\ \nonumber
&& - p\partial_{\theta}\delta Q + \gamma \partial_p
\left(p \delta Q + T \partial_p \delta Q\right),
\end{eqnarray}
which is to be compared to the corresponding Eq.~(\ref{eqself_b}) within the
Vlasov description (the linearised $x_b$ equation, being identical to
Eq.~(\ref{eqself_a}) is not reported). Using the Ans\"atze (\ref{ansatz}), we
finally find
\begin{eqnarray}
\label{FP_lambda}
\frac{i}{2} \bar{Y}^{\ast} {\cal E}_b \partial_p Q_0 &+&
  \left( \lambda - 2ip - \gamma \right) {\cal Q}\\
  && - \gamma \left(p \partial_p {\cal Q} +
  T\partial_p^2 {\cal Q} \right) = 0, \nonumber
\end{eqnarray}
which is to be solved together with the first of Eqs.~(\ref{FP_lambda}). In view
of its differential structure, it is doubtful whether a general analytic
solution can be found, and the development of a numerical analysis would be a
quite delicate task as well. Nevertheless, in the ``strong friction'' (SF)
limit, the atomic momentum can be adiabatically eliminated and one can thereby
perform a quite detailed investigation of the transition scenario. While a careful
discussion of the adiabatic elimination in the presence of noise can be found
in Ref.~\cite{FP84}, here we limit ourselves to presenting a sketchy but
substantially correct derivation where we set $\dot p = 0$ in Eq.~(\ref{FP_extern}).
By proceeding in this way, one obtains the Langevin equation
\begin{equation}
\label{Theta_reduce}
\dot \theta_{j} = \frac{1}{\gamma} \Im m(x_f x_b^{\ast}{\rm e}^{2i\theta_{j}}) +
\frac{\eta_{j}}{\gamma} ,
\end{equation}
that is equivalent to the one-variable Fokker-Planck (Smoluchowski) equation
\begin{equation}
\label{Smoluch}
\partial_t \rho + \frac{1}{\gamma} \partial_\theta
\Im m\left(x_f x_b^{\ast}{\rm e}^{2i\theta}\right) \rho - \frac{T}{\gamma}
 \partial_\theta^2 \rho = 0
\end{equation}
for the variable $\rho$
\begin{equation}
\label{rho_def}
\rho(\theta) = \int_{-\infty}^{\infty} Q(\theta,p)~dp.
\end{equation}
We remark that Eq.~(\ref{Theta_reduce}), which describes the dynamics of an
ensemble of mean field coupled oscillators, belongs to the class of Kuramoto
systems~\cite{Strogatz}. The main differences are: a Dirac-like distribution of
eigenfrequencies and a mean field self-consistently provided by a dynamical
equation.
By paralleling the approach in Section~\ref{Vlasov}, the stability of the
solution $x_f=\bar Y$, $x_b=0$, $\rho=1/2\pi$ can be studied by introducing the
infinitesimal perturbations $\delta x_b$ and $\delta \rho$. The linearisation of
Eq.~(\ref{Smoluch}) yields
\begin{equation}
\label{Smoluch_pert}
\partial_t \delta \rho = - \frac{1}{2\pi \gamma} \partial_\theta
\Im m\left(\bar Y \delta x_b^{\ast}{\rm e}^{2i\theta}\right) +
\frac{T}{\gamma} \partial_\theta^2 \delta \rho .
\end{equation}
Upon then inserting the Ans\"atze
$\delta \rho = r {\rm e}^{-2i \theta + \lambda t} + c.c$ and
$\delta x_b = {\cal E}_b {\rm e}^{\lambda t}$ into the above equation, one
obtains
\begin{equation}
\label{Solu1}
r=-\frac{{\cal E}_b \bar Y^{\ast}}{2\pi} \frac{1}{\lambda \gamma + 4T} ,
\end{equation}
and from the linearised equation for the field dynamics, Eq.~(\ref{eqself_a}),
\begin{equation}
\label{Lambda_eq}
\lambda = -(1+i \bar \Delta) + i\frac{C |\bar Y|^2}{\lambda \gamma + 4 T},
\end{equation}
where now,
\begin{equation}
\label{Bunch_proba_reduce}
B_{\pm 1} = \int_{0}^{2\pi} e^{\pm 2i\theta} \rho(\theta)~d\theta .
\end{equation}

The threshold can be now obtained by setting $\lambda = -i \beta$ (where $\beta$
is a real number) in Eqs.~(\ref{Solu1},\ref{Lambda_eq}). The resulting two real
equations for $\beta$ and $|\bar Y|$ are
\begin{eqnarray}
\label{syst_general_a}
\gamma \beta^2 - \bar \Delta \gamma \beta - 4 T &=& 0,\\
\label{syst_general_b}
-\beta \left(\gamma + 4T\right) + 4 \bar \Delta T &=& C |\bar Y|^2.
\end{eqnarray}
Only the negative $\beta$ solution of Eq.~(\ref{syst_general_a}) leads to a
physically acceptable solution $|\bar Y|^2 > 0$. The threshold equations finally
read
\begin{eqnarray}
\label{Thres_a}
\beta_{TH} &=& \frac{\bar \Delta}{2} - \sqrt{\left(\frac{\bar \Delta}{2}\right)^2 +
  4 \frac{T}{\gamma}}, \\
\label{Thres_b}
|\bar Y_{TH}|^2 &=& \frac{\bar \Delta}{2C} (4T - \gamma) + \frac{1}{C}(4T +
\gamma) \sqrt{\left(\frac{\bar \Delta}{2}\right)^2
  + 4 \frac{T}{\gamma}}  \nonumber .
\end{eqnarray}
The first equation shows that the probe field is {\it always} red detuned at
threshold. This result is in agreement with the outcome of the Vlasov model
(see Section~\ref{Vlasov}) even though these two models correspond to different
thermalisation mechanisms.

In the case of a resonant cavity field, $\bar \Delta = 0$, the above
equations reduce to
\begin{eqnarray}
\label{syst_general}
\beta_{TH} &=& - 2 \sqrt{\frac{T}{\gamma}}, \\
|\bar Y_{TH}|^2 &=& \frac{2}{C}(4T +\gamma)\sqrt{\frac{T}{\gamma}}.
\end{eqnarray}
In the SF limit, one can not only determine analytically the threshold
condition, but can also go beyond, describing the regime above threshold.
By inserting the expansion of $\rho$,
\begin{equation}
\label{decomp}
\rho \left(\theta,t\right) = \sum_{n=-\infty}^\infty f_n(t){\rm
e}^{in(2\theta+\beta t)} ,
\end{equation}
into Eq.~(\ref{Smoluch}), one obtains the recurrence relation
\begin{equation}
\label{Rec_rel}
 \gamma \dot{f}_{n} = - in\gamma \beta f_{n} - n\left[R f_{n-1} -
 R^{\ast}f_{n+1}\right] - 4 T n^{2}f_{n} \, ,
\end{equation}
where we have set $x_f x_b^{\ast}=R {\rm e}^{i \beta t}$, $R$ being a complex
constant. For $n>0$, the stationary state is given by
\begin{equation}
\label{station_f}
\left(4 T n + i\gamma \beta \right) f_{n} + \left[R f_{n-1} -
R^{\ast}f_{n+1}\right]=0 \, .
\end{equation}
By introducing $Z_{n}=R^{\ast} (f_{n+1}/ f_{n})$ and solving the equation by
means of the continuous fraction method, we obtain the recurrence relation
\begin{equation}
\label{Frac_cont}
Z_{n-1}=\frac{|R|^2}{Z_n -\left(4T n+i \gamma \beta\right)}.
\end{equation}
The normalisation condition for $\rho$ gives $2\pi f_0 = 1$, while the atomic
contribution for $x_b$ and $x_f$ can be then determined from
Eq.~(\ref{Bunch_proba_reduce}),
\begin{eqnarray}
\begin{array}{c c c c c}
B_1&=& 2\pi f_{-1}{\rm e}^{-i\beta t}&=&\frac{Z_0^{\ast}}{R}{\rm e}^{-i\beta t} \\
&&&&\\
B_{-1} &=& 2\pi f_{1}{\rm e}^{i\beta t}&=&\frac{Z_0}{R^{\ast}}{\rm e}^{i\beta t} \, .\\
\end{array}
\end{eqnarray}
This methods enables us to go beyond the linear analysis determing the
dependence of {\it both} $x_f$ and $x_b$ on the injected field (see
Section~\ref{predictions}), with the only warning that $Z_0$ has to be computed
numerically.

\section{Predictions}\label{predictions}

In this section we discuss the results derived in Section~\ref{Large} in the
large detuning limit for both the Fokker-Planck (see Sec.~\ref{FP}) and Vlasov
(Sec.~\ref{Vlasov}) models. We first test the validity of the two approaches by
comparing them to the experimental results of Ref.~\cite{KCZC03}.
The experiment has been performed with cold $^{85}\text{Rb}$ atoms in a high-$Q$
ring cavity. The atomic parameters are: $m=1.4~10^{-29}$kg,
$\mathcal{D} = 1.5~10^{-29}\text{Cm}^{-1}$ with relaxation rates for atomic
polarisation and inversion $2\gamma_{\perp} = \gamma_\parallel = 5.9$MHz, respectively.
Moreover, the atomic frequency is $\omega = 3.77 \times 10^{14}$Hz (for the $D_1$ line),
while the input field is shifted 1THz away from it.
The beam diameter is $130\mu$m. The cavity linewidth is $22$kHz,
the medium's length $L=10^{-3}m$, and the transmissivity ${\mathcal T}=1.8 \times 10^{-6}$. Since a
servo-mechanism continuously adapts the dressed cavity resonance to the
frequency of the injected field, we assume $\bar \Delta=0$.

The equilibrium gas temperature, in the absence of collective interaction, is
experimentally known with a large uncertainty.  We use an estimated value,
$\bar T \approx 250\mu$K, which lies within the range of uncertainty.  The
second crucial parameter which caracterizes the thermalisation is the damping
constant, experimentally evaluated in Ref.~\cite{KCZC03} to be $\gamma = 9$.
We will thus use $\gamma = \gamma_c = 9$.
\begin{figure}[ht!]
\includegraphics[clip,width=8.cm]{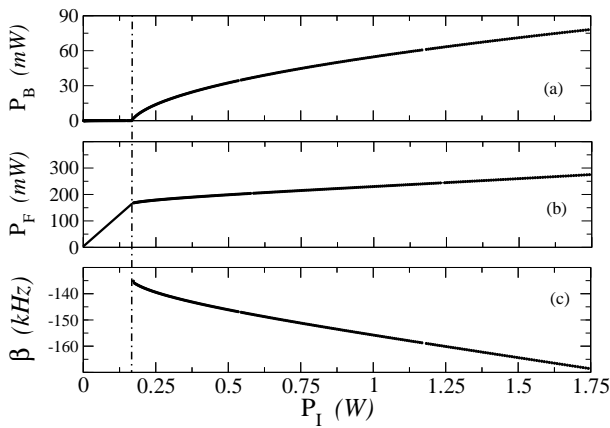}
\caption{Bifurcation diagram numerically computed using the Fokker-Planck
model.  Figs. (a), (b), (c), present respectively the backward field power,
$P_B$, the forward field power, $P_F$ and the detuning between backward and
forward field, $\beta$, as function of the injected power $P_I$. All quantities
are in physical units.  The parameters correspond to those of the experiment
\cite{KCZC03}.}
\label{FP_thres}
\end{figure}

We first report the results of the Fokker-Planck model since it describes more
appropriately the molasses dynamics. In Fig.~\ref{FP_thres}a
one remarks, from the shape of the backward field power, the existence of a 
continuous phase transition. The threshold is located at  $P_I \approx$ 0.2W,
compatible with the measurements of Ref.~\cite{KCZC03}, which
reports backward lasing as occurring for a few Watt of injected power.

The possible underestimate in the threshold value, which we obtain from our
approach, can be attributed to two experimental features that cannot be 
taken into account in a simple way.  First, the servo-control which keeps the
cavity on resonance by {\it maximizing the forward transmitted
field}~\cite{KCZC03} cannot be correctly modeled by simply setting
$\overline{\Delta}=0$. Indeed, its operation amounts to a dynamical process
which counteracts the birth of the backward emission, since the latter removes
energy from the forward field.  As a result, one expects the threshold to be
displaced towards higher input power values.  Second, the spatial modulation
imprinted onto the sample by the molasses beam, with a wavelength that differs
from that of the collective process (by a factor close to $\sqrt{2}$),
introduces a competition between two incompatible length scales.  It is likely
that this effect should also contribute to raising the critical input power even
further.

The detuning value, $\beta$, at which the backward field arises (cf.
Fig.~\ref{FP_thres}c) also agrees with the observations~\cite{KCZC03}, which
report a red-shift around 100kHz.  Moreover, the continued fraction expansion
discussed in the previous section also allows for a determination of the
behavior of the forward field as a function of the input power (cf.
Fig.~\ref{FP_thres}b). Below threshold, $P_F$ grows linearly as a function of
the injected power, $P_I$, as no energy is transferred to the backward field.
Above threshold, a fraction of the injected field feeds the backward mode, thus
leading to a slower growth of $P_F$.
\begin{figure}[ht!]
\includegraphics[clip,width=8.cm]{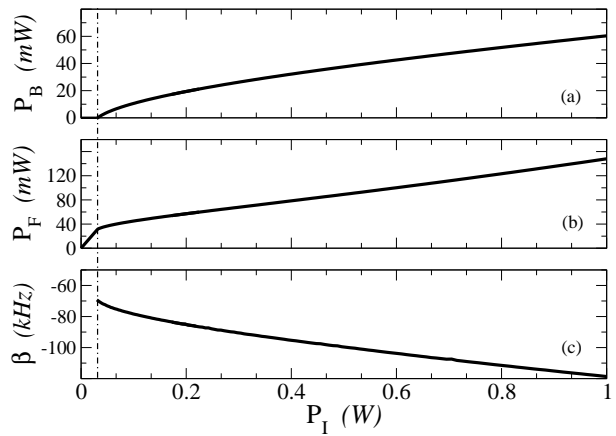}
\caption{Bifurcation diagram numerically computed using the Vlasov model.
Figs. (a), (b), (c), present respectively the backward field power, $P_B$, the
forward field power, $P_F$, and the detuning between backward and forward
field, $\beta$, as function of the injected power $P_I$.  All quantities
are in physical units.  The parameters correspond to those of the experiment
~\cite{KCZC03}.}
\label{Vlasov_Thres}
\end{figure}

Although one does not expect the Vlasov description to closely reproduce the
experimental results reported in Ref.~\cite{KCZC03}, it is worth analyzing its
predictions; this will help us clarifying the role of the relaxation mechanisms and
will sharpen our understanding of the problem.  From Fig.~\ref{Vlasov_Thres}
one sees that, for the same parameter values used in the Fokker-Planck approach,
the threshold is somewhat lower and the red shift in the backward field
frequency is smaller as well.  Nevertheless, the differences are not so
dramatic, since, while the threshold occurs for somewhat lower values of the
input power, the frequency shift shows an even somewhat better agreement with
the observations~\cite{KCZC03}.  In addition, the uncertainty on some of the
experimental parameters is large enough to render a sure discrimination between
the modeling based on a Vlasov or on a Fokker-Planck process quite difficult. 
Because of the intrinsic paradox, given by the inclusion in the experiment
of the optical molasses which ought to exclude the Vlasov description, this result
strongly indicates that the current state of the system's characterization, both
experimental and theoretical, is not adequate; hence, a large amount of caution
in drawing conclusions ought to be exercised.
\begin{figure}[ht!]
\includegraphics[clip,width=8.cm]{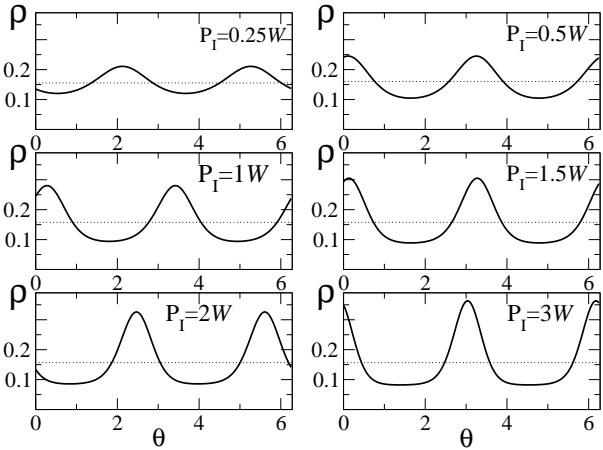}
\caption{Position distribution $\rho(\theta)$, for different values of the
injected power $P_I$, in physical units. In each part of the figure, the dotted
line corresponds to the uniform distribution. Parameters correspond to the
experimental values of Ref.~\cite{KCZC03}.}
\label{FP_bunch}
\end{figure}

In addition to the direct comparison to the experiment, we can also examine some
specific predictions coming from each model.  The distribution of atomic
positions $\rho$ predicted by the Fokker-Planck model is plotted in
Fig.~\ref{FP_bunch}: the different parts of the figure correspond to increasing values of
the injected power $P_I$ (the other parameters remaining unchanged). One can
clearly see the rise of the modulation with an increasing role played by higher
harmonics.
\begin{figure}
\includegraphics[width=5.cm,angle=270]{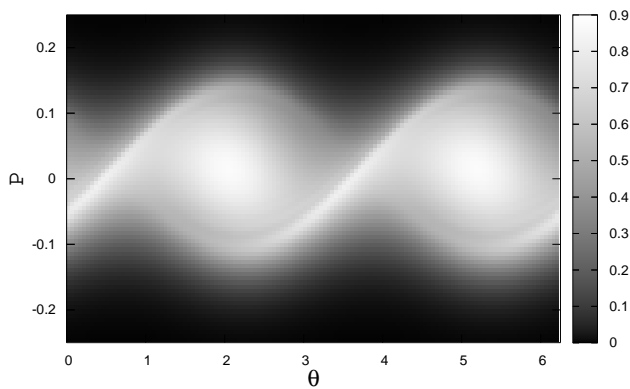}
\caption{Joint distribution $Q(\theta,p)$, above threshold. The parameter values
are the same as those of the experiment~\cite{KCZC03} with input
power of 480mW.}
\label{Dist_Vlasov_2D}
\end{figure}

\begin{figure}[ht!]
\includegraphics[clip,width=7.cm]{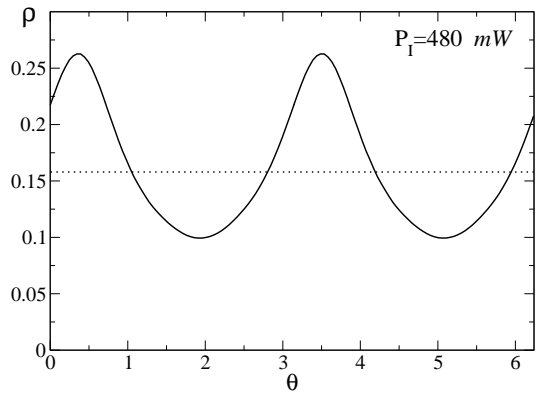}
\caption{Atomic position distribution $\rho(\theta)$, as function of $\theta$,
above threshold. The dotted line corresponds to the uniform distribution, below
threshold.}
\label{Dist_Vlasov_1D}
\end{figure}

In the Vlasov description, there is no equivalent regime characterized
exclusively by the distribution of $\theta$-values. The probability density 
$Q(\theta,p)$
for an input intensity above threshold is plotted in Fig.~\ref{Dist_Vlasov_2D},
where one can indeed appreciate the need to account for both the $p$ and $\theta$
dependence in this framework. Nevertheless, one can compute  $\rho(\theta)$ by
integrating $Q(\theta,p)$ (see Eq.~(\ref{rho_def})). The results, plotted in
Fig.~\ref{Dist_Vlasov_1D}, reveal again a qualitative agreement with
the Fokker-Planck approach. The larger modulation for the same input power
is a consequence of the smaller value of the threshold in the Vlasov
description.

\section{Conclusions and perspectives}\label{concl}

In this paper we have presented a general framework for the description of the
{\it spontaneous} appearence of a {\it stationary} density grating arising from the
interaction between an atomic sample enclosed in a bidirectional ring cavity
and an injected, quasi-resonant field.  This contribution completes the picture
previously sketched out in Refs.~\cite{PLP01,JLP03}, where we have shown
that the inclusion of a relaxation mechanism for the external degrees of
freedom is indispensable for a correct modeling of the interaction, and is
responsible for the appearence of a stationary backward field. In the previous
work, we focused on the existence of a different kind of phase transition,
initiated by a grating in the phase of the atomic polarization.  Here, with the
help of a detailed numerical analysis we have proven beyond doubt, that a true
stationary CARL action -- i.e., initiated by the birth of an atomic density
modulation -- can spontaneously occur.  For the investigation, we have chosen
to study a bidirectional ring cavity model~\cite{YN01,PYN02} because of the
fuller and more correct description that the latter provides of a real system.
In the course of the presentation, we have shown that in certain limits the
ring cavity model reduces to the single-pass one~\cite{carl,PLP01,JLP03}; this
legitimizes, a posteriori, the use of the latter approach under such
conditions.

Following a discussion of the main relaxation mechanisms that one may expect to
find in an experiment, we have offered an analytical description of the two
most likely ones on the basis of a probabilistic description, in the large
detuning limit. Comparison between our predictions and recent experimental
results~\cite{KCZC03} shows a semi-quantitative agreement, surprisingly, for
both the Fokker-Planck and the  Vlasov approach.  Besides the need for a more
accurate experimental characterization of the phenomenon, including a better
determination of the parameter values, and systematic measurements of the
threshold behavior and of the functional dependence of the backward field power
and frequency, additional theoretical work needs to be completed. In particular,
a more accurate modeling of the relaxation mechanism represented by the optical
molasses needs to be achieved, together with an analysis of the limits of
validity of the descriptions proposed for the losses. Finally, it will be useful
to investigate the transtion scenario when two mechanisms separately described
here and in Refs.~(\cite{YN01,PYN02}) cooperate to induce a backward field;
this is likely to occur in the range of smaller detunings when the atomic
variables cannot be eliminated.

\acknowledgments
The authors warmly thank Ph.W. Courteille for sharing information on the
experiment before publication. MP wishes to acknowledge INOA for financial
support, L.M. Narducci and M. Le Bellac for useful discussions. This work has
been partial funded by the FIRB-contract n.~RBNE01CW3M\_001.


\begin{thebibliography}{10}
\bibitem{com1} For instance, the original model \protect\cite{carl}, derived under
the approximation of a constant input field, does not conserve energy even
in the absence of dissipation.
\bibitem{note1} While the atomic polarization and population can be safely
eliminated by following the procedure described in Ref.~\cite{JLP03}, more
questionable is the treatment of position and momentum.  However, preliminary
investigations indicate that this is possible at least in some parameter
ranges.
\bibitem{note3} The net effect of this randomization is to increase the threshold
value of the injected field, because of the disorder introduced.
\bibitem{note4} An analogous stability analysis was carried out also in
Ref.~\cite{BV96}, where, however, the possible occurence of a phase
transition was missed because of the additional presence of a probe field.
In fact, as is the case of ferromagnetic systems, the presence of an external
field destroys the second-order transition.
\end{thebibliography}
\end{document}